\documentclass[aps,prd,nofootinbib,twocolumn,superscriptaddress,preprintnumbers,balancelastpage,longbibliography]{revtex4-2}

\usepackage{amsmath}
\usepackage{orcidlink}
\usepackage{graphicx}
\usepackage{dcolumn}
\usepackage{bm}

\usepackage{xcolor}

\def\lsim{\mathrel{\raise.3ex\hbox{$<$\kern-.75em\lower1ex\hbox{$\sim$}}}}
\def\gsim{\mathrel{\raise.3ex\hbox{$>$\kern-.75em\lower1ex\hbox{$\sim$}}}}

\newcommand{\ord}[1]{\mathcal{O}{(#1)}}
\newcommand{\beq}{\begin{equation}}
\newcommand{\eeq}
{\end{equation}}
\newcommand{\bea}{\begin{eqnarray}}
\newcommand{\eea}{\end{eqnarray}}
\newcommand{\eps}{\varepsilon}

\begin{document}

\title{Novel Signatures of Matter-Induced Dark Matter Decay\\ in Large-Volume Neutrino Telescopes}

\author{Hooman Davoudiasl}
\email{hooman@bnl.gov}
\affiliation{High Energy Theory Group, Physics Department,
Brookhaven National Laboratory, Upton, NY 11973, USA}

\author{Dan Hooper}
    \email{dwhooper@wisc.edu}
\affiliation{
    Department of Physics, Wisconsin IceCube Particle Astrophysics Center, University of Wisconsin, Madison, WI 53706, USA
}

\author{Samyak Jain}
\email{samyak@icecube.wisc.edu}
\affiliation{
    Department of Physics, Wisconsin IceCube Particle Astrophysics Center, University of Wisconsin, Madison, WI 53706, USA
}

\date{\today}

\begin{abstract}

Large-volume neutrino telescopes offer a unique opportunity to search for decaying dark matter through events containing a pair of energetic, highly non-collimated muon tracks emerging from a common vertex. Such events would have negligible Standard Model backgrounds and would constitute a striking signature of new physics. Conventional dark matter annihilation or decay, however, is too strongly constrained to produce an observable rate of such events. We therefore consider scenarios in which an excited dark matter state is extremely long-lived in vacuum but decays much more rapidly in the presence of ordinary  matter. We present two realizations of this mechanism. In the first, a long-range scalar field sourced by ordinary matter modifies the dark-sector mass spectrum, kinematically opening the decay $\chi_2 \rightarrow \chi_1 Z'$ near the Earth while leaving it forbidden in vacuum. In the second, the scalar background induces kinetic mixing between a heavy $Z'$ and the photon, greatly enhancing the three-body decay $\chi_2\to\chi_1\mu^+\mu^-$ in matter-rich environments. We calculate the resulting distributions of muon energies and opening angles and show that viable regions of parameter space can yield observable event rates in IceCube, KM3NeT, and other large-volume neutrino telescopes while remaining consistent with existing constraints. We also briefly consider the sensitivity of IceCube to multi-muon events produced by the decays of cosmologically long-lived charged particles with masses $\gtrsim 1$ TeV.

\end{abstract}

\maketitle


\section{Introduction} \label{introduction}

Neutrino telescopes provide a unique means to test and constrain the nature of dark matter. Searches for dark matter annihilation or decay in the Sun \cite{ ADRIANMARTINEZ201669, IceCube:2021xzo}, the Earth \cite{icecube_dark_matter_earth}, and the Galactic Halo \cite{IceCube:2023ies,IceCube:2015rnn, IceCube:2017rdn, ANTARES:2020leh, ANTARES:2019svn, KM3NeT:2024xca,IceCube:2025fcn,IceCube:2026rbh} have yielded significant constraints that are expected to improve with future experiments~\cite{qj8m-v4mr, Bell:2020rkw}. Such searches, however, are subject to significant backgrounds from atmospheric neutrinos and muons and are often not competitive with constraints from other direct~\cite{LZ:2024zvo, XENON:2025vwd, PandaX:2024qfu, CRESST:2019jnq, PICO:2019vsc} and indirect~\cite{Planck:2018vyg,John:2021ugy,Blanco:2018esa,Circiello:2026inp,DelaTorreLuque:2024ozf} detection experiments.

In this paper, we consider a new and potentially powerful signal of decaying dark matter in large-volume neutrino telescopes. In particular, we consider dark matter particles that can produce a pair of energetic, highly non-collimated muons within the detector volume. Such events would have essentially no Standard Model backgrounds and would constitute a smoking-gun signature of new physics. We focus on scenarios in which an excited dark matter state is extremely long-lived in vacuum but decays much more rapidly in the presence of ordinary matter. We motivate searches for the resulting dimuon events and estimate the sensitivities of large-volume neutrino telescopes to this class of models.  

While not the main focus of this work, we also consider the possibility that multi-muon signals could arise from a small population of cosmologically long-lived charged massive particles (CHAMPs) \cite{Cahn:1980ss,DeRujula:1989fe,Dimopoulos:1989hk} with masses above the TeV scale that have accumulated within the Earth. We do not assume any specific origin or production mechanism for these particles, but show that interesting bounds on their lifetime can be obtained from IceCube data. For recent work on other methods of constraining CHAMPs, see Ref.~\cite{Ebadi:2026umu}.

The remainder of this paper is structured as follows. In Sec.~\ref{sec:prelude}, we consider conventional dark matter annihilation and decay and show that these processes cannot produce detectable rates of dimuon events. In Sec.~\ref{DMdecay}, we introduce scenarios in which dark matter decays preferentially in the presence of ordinary matter, leading to potentially observable rates of high-energy dimuon events in large-volume neutrino telescopes. We accomplish this by introducing a long-range interaction through which visible sector matter ({\it i.e.} nucleons) sources a classical scalar background. This background either modifies the dark-sector mass spectrum or induces couplings required for the decay, allowing the dark matter state to be extremely long-lived in vacuum and yet able to decay far more efficiently within or near the Earth. In Sec.~\ref{sec:kinematics}, we describe the kinematics of these decays, and in Sec.~\ref{sec:NS} we discuss constraints from neutron-star heating. In Sec.~\ref{sec:CHAMPs}, we investigate the prospects for detecting multi-muon signals from the decay of CHAMPs. We summarize our results and conclusions in Sec.~\ref{sec:summary}.

\section{Prelude: Conventional Dark Matter Annihilation and Decay}
\label{sec:prelude}

As a first example, consider a dark matter candidate, $\chi$, which annihilates into a muon-antimuon pair with a velocity-averaged cross section, $\langle \sigma v \rangle_{\chi \chi}$. Assuming that $\chi$ is its own antiparticle, the corresponding annihilation rate per unit volume is given by
\begin{align}
&\Gamma_{\chi \chi \rightarrow \mu^+ \mu^-} \approx  \frac{n^2_\chi(r_{\odot}) \langle \sigma v \rangle_{\chi \chi}}{2} \\
& \,\,\,\,\,\, \approx 6 \times 10^{-11} \,{\rm yr}^{-1} \, {\rm km}^{-3} \, \bigg( \frac{{\rm TeV}}{m_\chi}\bigg)^2 \bigg( \frac{\langle \sigma v \rangle_{\chi \chi}}{2.2\times 10^{-26} \, {\rm cm}^3/{\rm s}}\bigg),\nonumber 
\end{align}
where we have adopted a local dark matter density of $\rho_{\chi}(r_{\odot}) = 0.4 \, {\rm GeV/cm}^3$~\cite{Pato:2015dua} and assumed a branching fraction of unity to $\mu^+ \mu^-$. Since Galactic dark matter is nonrelativistic, the two muons produced in such an event would be approximately back-to-back in the detector frame. 

Alternatively, the rate of such events produced by a dark matter particle that decays to muons with a lifetime, $\tau_{\chi}$, is given by:
\begin{align}
&\Gamma_{\chi \rightarrow \mu^+ \mu^-} \approx \frac{n_\chi(r_{\odot})}{\tau_\chi} \\
& \,\,\,\,\,\, \approx 1.3 \times 10^{-8} \,{\rm yr}^{-1} \, {\rm km}^{-3} \, \bigg( \frac{{\rm TeV}}{m_\chi}\bigg) \bigg( \frac{10^{27} \text{ sec}}{\tau_\chi}\bigg),\nonumber 
\end{align}
where again we have assumed a branching fraction of unity to $\mu^+ \mu^-$.

If such an annihilation or decay occurred within the volume of a suitable detector, it would produce two approximately back-to-back muon tracks emerging from a common vertex. This topology would have negligible Standard Model background and would constitute a highly distinctive signature of new physics. The calculations presented above, however, demonstrate that conventional dark matter annihilation or decay cannot produce an observable number of such events. For annihilation cross sections or decay lifetimes consistent with constraints from the cosmic-ray positron spectrum~\cite{Bergstrom:2013jra,Ibarra:2013zia,Cavasonza:2016qem,John:2021ugy} and the isotropic gamma-ray background~\cite{DiMauro:2015tfa,Blanco:2018esa}, the rates of dimuon events are expected to be undetectably small, even for kilometer-scale detectors.

For dark matter to produce muon pairs at a rate that could be detectable in large-volume neutrino detectors, we will consider models in which the dark matter decays preferentially in the presence of baryonic matter. Such phenomenology could be realized in a variety of ways and, in the following section, we will outline two such possibilities.

\section{Matter-Induced Dark Matter Decays}\label{DMdecay}

In each of the scenarios described in this section, a fraction of the dark matter consists of a state, $\chi_2$, that is extremely long-lived in vacuum but can decay much more rapidly in regions near large concentrations of visible matter. These decays produce a lighter dark-sector state, $\chi_1$, and a pair of muons:
\begin{align}
\label{eq:process}
\chi_2 &\rightarrow \chi_1 + Z'^{(\star)} \rightarrow \chi_1 + \mu^+ \mu^-,
\end{align}
where the $Z'$ is a new gauge boson that may be either on shell or virtual. Such a $Z'$ could arise, for example, in models featuring a spontaneously broken
$U(1)_{L_\mu-L_\tau}$ symmetry~\cite{He:1990pn,He:1991qd,Heeck:2011wj,Altmannshofer:2014pba,Escudero:2019gzq}.

Decays of this kind would produce dimuon events in a large-volume neutrino detector at the following rate per unit volume:
\begin{align}
\frac{d N_{\mu \mu}}{dV dt}  &= \frac{n_{\chi_2} B_{\mu}}{\tau_{\chi_2}} \\
&=\frac{f_{\chi_2} \rho_{\rm DM} B_{\mu}}{m_{\chi_2} \tau_{\chi_2}} \nonumber \\
&\approx 0.6 \, {\rm yr}^{-1} \, {\rm km}^{-3} \,\bigg(\frac{f_{\chi_2}}{0.5}\bigg)\bigg(\frac{B_{\mu}}{1}\bigg)\bigg(\frac{1 \, {\rm TeV}}{m_{\chi_2}}\bigg)\bigg(\frac{10^{19} \,{\rm s}}{\tau_{\chi_2}}\bigg), \nonumber
\end{align}
where $f_{\chi_2}$ is the fraction of the dark matter in the form of $\chi_2$, $\rho_{\rm DM} \approx 0.4 \, {\rm GeV/cm}^3$ is the local dark matter density, $\tau_{\chi_2}$ is the total lifetime of $\chi_2$ in the environment of the detector, and $B_{\mu}$ is the branching fraction of $\chi_2$ decays that produce a $\mu^+ \mu^-$ pair. 

To evade the stringent constraints on conventional decaying dark matter, the decay rate must be highly suppressed in vacuum while becoming substantially enhanced near sufficiently large concentrations of ordinary matter. 
 In the following subsections, we present two illustrative mechanisms that could realize this behavior.

\subsection{Matter-Induced Mass Shifts}
\label{sec:massshifts}

For $\chi_2$ to be extremely long-lived while being able to decay near sufficiently large concentrations of ordinary matter, we consider scenarios in which ordinary matter sources a classical scalar background that modifies the dark-sector mass spectrum. This is similar in spirit to models with environmentally dependent particle masses, including chameleon and symmetron constructions~\cite{Khoury:2003aq,Khoury:2003rn,Hinterbichler:2010es,Olive:2007aj}.

As an example, consider a model with two fermionic dark-sector states, $\chi_1$ and $\chi_2$, a light scalar, $\phi$, and a massive vector boson, $Z'$, with the following low-energy Lagrangian:
\begin{align}
\mathcal{L} \supset &
\sum_{i=1,2} \bar{\chi}_i \left( i \gamma^{\mu} \partial_{\mu}-m_{\chi_i} \right)\chi_i
+\frac{1}{2}\partial_\mu \phi \partial^{\mu} \phi -V(\phi) \nonumber \\
&-\frac{1}{4}Z'_{\mu\nu}Z'^{\mu\nu} 
+\frac{1}{2}m_{Z'}^2 Z'_\mu Z'^\mu \nonumber \\
&-y_1\phi \bar{\chi}_1\chi_1
-y_2\phi \bar{\chi}_2 \chi_2
-y_N\phi \bar N N
\nonumber \\
&+ (g_\chi Z'_\mu \bar{\chi}_1 \gamma^\mu \chi_2
+{\rm h.c.})
+g_\mu Z'_\mu \bar{\mu} \gamma^{\mu} \mu, 
\label{eq}
\end{align}
where $Z_{\mu \nu}' \equiv \partial_{\mu} Z_{\nu}' - \partial_{\nu} Z_{\mu}'$, $V(\phi)$ is the scalar potential, $m_Z'$ is the mass of the $Z'$, $y_1$, $y_2$ and $y_N$ are Yukawa couplings, and $g_{\chi}$ and $g_{\mu}$ are vector couplings. The $-y_N \phi \bar{N} N$ term in this expression allows ordinary matter ($N$ represents a nucleon) to source a classical scalar field. The terms of the form $-y_i \phi \bar{\chi_i} \chi_i$, in turn, make the effective dark matter masses dependent on that field. 

Treating the Earth as an approximately uniform spherical body of radius $R_{\oplus}\approx 6400$~km  containing $M_{\oplus}/m_N\sim 10^{51}$ total nucleons, the field near its surface (in the long-range limit, $m_\phi^{-1} \gg R_{\oplus }$) is parametrically given by
\begin{align}
\label{eq:potential}
\phi(R_{\oplus})\approx - \frac{y_N M_{\oplus}} {4\pi R_{\oplus} m_N}.
\end{align}
More general potentials and matter couplings lead to nonlinear profiles and can exhibit thin-shell or other screening effects~\cite{Khoury:2003aq,Khoury:2003rn,Hinterbichler:2010es}. Eq.~(\ref{eq:potential}) should therefore be regarded as an illustrative unscreened limit. Note that for simplicity, we will focus on scenarios in which $y_2 =0$. This also ensures that $\chi_2$ particles will not be deflected by the potential before reaching the detector.  However, we will keep the following discussion general. 

The scalar background changes the effective masses of the two dark-sector states,
\begin{align}
m_{\chi_i}^{\rm eff}(\mathbf{x})=m_{\chi_i}+y_i\phi(\mathbf{x}),
\label{eq:scalar-Yukawa}
\end{align}
where $m_{\chi_2}$ and $m_{\chi_1}$ are the masses in vacuum and $\phi(\mathbf{x})$ is the value of the classical scalar field sourced by ordinary matter. This changes the effective splitting between these states,
\begin{align}
m_{\chi_2}^{\rm eff}(\mathbf{x})
-
m_{\chi_1}^{\rm eff}(\mathbf{x})
=
m_{\chi_2}-m_{\chi_1}
+
\left(y_2-y_1\right)\phi(\mathbf{x}).
\end{align}

For $m_{\chi_2}-m_{\chi_1} < m_{Z'}$, the two-body decay, $\chi_2 \rightarrow \chi_1 +Z'$, is kinematically forbidden in vacuum, while the off-shell three-body decay, $\chi_2 \rightarrow \chi_1 Z'^{\star} \rightarrow \chi_1 \mu^+ \mu^-$, can be highly suppressed if $g_{\mu}$ is small. Near the Earth, however, the scalar background can raise the effective mass splitting above threshold, provided that $\left(y_2-y_1\right)\phi(\mathbf{x})> m_{Z'}-(m_{\chi_2}-m_{\chi_1})$. For $y_2=0$ and $y_1 y_N > 0$, the Earth-sourced scalar field lowers $m_{\chi_1}^{\rm eff}$, thereby increasing the effective mass splitting. Importantly, this environmental mass shift need not be comparable to the full mass splitting; it need only bridge the smaller gap of $\sim m_{Z'} - (m_{\chi_2} - m_{\chi_1})$. Due to the very large number of nucleons present in the Earth, sizable shifts in the $\chi_{1,2}$ masses can result even from very small values of the respective Yukawa couplings:
\begin{align}
|\Delta m_{\chi_i}| &= |y_i \phi(\mathbf{x}) | \\
&\sim 100 \,  {\rm GeV} \times \bigg|\frac{y_i y_N}{10^{-26}}\bigg|.\nonumber
\end{align}

The width for this on-shell decay process is given by 
\begin{align}
\Gamma(\chi_2 &\rightarrow \chi_1 Z') \\
&=  \frac{g_{\chi}^2 \lambda^{1/2}((m^{\rm eff}_{\chi_2})^2, (m^{\rm eff}_{\chi_1})^2,m^2_{Z'})}{16 \pi (m^{\rm eff}_{\chi_2})^3 m_{Z'}^2}  \, \Theta(m^{\rm eff}_{\chi_2}-m^{\rm eff}_{\chi_1}-m_{Z'})\nonumber \\
&\,\,\,\, \times ((m^{\rm eff}_{\chi_2}-m^{\rm eff}_{\chi_1})^2-m^2_{Z'})((m^{\rm eff}_{\chi_2}+m^{\rm eff}_{\chi_1})^2+2m^2_{Z'}), \nonumber
\end{align}
where $\lambda(a,b,c) = a^2+b^2+c^2-2ab-2ac-2bc$. In the heavy, small-splitting limit, $m^{\rm eff}_{\chi_2},m^{\rm eff}_{\chi_1} \gg (m^{\rm eff}_{\chi_2}-m^{\rm eff}_{\chi_1}), m_{Z'}$, this reduces to
\begin{align}
\Gamma(\chi_2 \rightarrow \chi_1 Z') &\approx  \frac{g_{\chi}^2}{2 \pi m_{Z'}^2}  \,  [(m^{\rm eff}_{\chi_2}-m^{\rm eff}_{\chi_1})^2-m^2_{Z'}]^{3/2} \nonumber \\ 
&\,\,\,\,\,\,\,\,\,\,\,\, \times \Theta(m^{\rm eff}_{\chi_2}-m^{\rm eff}_{\chi_1}-m_{Z'}).
\end{align}
Above threshold, this corresponds to a lifetime of
\begin{align}
\tau_{\chi_2 \rightarrow \chi_1 Z'} &= \Gamma(\chi_2 \rightarrow \chi_1 Z')^{-1} \\
&= \frac{2 \pi m_{Z'}^2} {g_{\chi}^2 [(m^{\rm eff}_{\chi_2}-m^{\rm eff}_{\chi_1})^2-m^2_{Z'}]^{3/2}}  \nonumber \\
&\approx 1.5 \times 10^{18} \, {\rm s} \times \bigg(\frac{10^{-22}}{g_{\chi}}\bigg)^2\bigg(\frac{m_{Z'}}{200 \, {\rm GeV}}\bigg)^2 \nonumber \\
&\,\,\,\,\,\,\,\,\,\, \times \bigg(\frac{(300 \, {\rm GeV})^2-(200 \, {\rm GeV})^2}{[(m^{\rm eff}_{\chi_2}-m^{\rm eff}_{\chi_1})^2-m^2_{Z'}]}\bigg)^{3/2}. \nonumber
\end{align}
Thus an observable rate of muon pair events could be generated even for extremely small values of the dark gauge coupling, $g_{\chi}$.

\subsection{Matter-Induced Kinetic Mixing}
\label{sec:couplingshifts}

As a second possibility, we consider a scenario in which the $Z'$ is heavy and thus must be off-shell:
\beq
\chi_2 \to \chi_1\, Z'^{*}\to \chi_1 \,\mu^+\mu^-\,.
\label{eq:off-shell-process}
\eeq
We adopt a model in which the coupling of the $Z'$ to muons is highly suppressed in vacuum but becomes enhanced in regions near large densities of ordinary  matter.  To realize this behavior, we introduce a long-range scalar, not to change the mass of $\chi_1$, but instead to make the kinetic mixing between the $Z'$ and the photon environmentally dependent~\cite{Davoudiasl:2022ubh}:
\beq
\mathcal{L} \supset \frac{1}{2}\left(\frac{\phi}{M}\right)^n
Z_{\mu\nu}' F^{\mu\nu}\,,
\label{eq:kin-mix}
\eeq
where $Z'_{\mu\nu}$ and $F_{\mu\nu}$ are the field-strength tensors of the $Z'$ and photon, respectively, and $n$ is an integer $n\geq 1$, which depends on the ultraviolet physics.  We will adopt $n=2$ going forward, as an illustrative example.

Using Eq.~(\ref{eq:potential}), the value of the matter-induced scalar field is given by
\begin{align}
\phi(R_\oplus) \sim 9\times 10^3~\text{GeV}\, \times \bigg(\frac{y_N}{10^{-24}}\bigg).
\label{eq:E-potential}
\end{align}

Far from dense astronomical bodies, the matter-sourced scalar background is extremely small. For example, taking the mean number density of nucleons in the interstellar medium to be $\sim 1$~cm$^{-3}$ and the scalar range to be $\sim R_\oplus\sim 10^4$~km yields $\phi\sim 10^{-20}$~GeV for $y_N\sim 10^{-24}$, near the limit from fifth-force constraints \cite{Fayet:2017pdp,MICROSCOPE:2022doy}.  Hence, we can safely ignore any kinetic mixing induced by $\phi$ far away from dense astronomical bodies.

Let us define $\eps\equiv \phi^2/M^2$ and take the interaction (\ref{eq:kin-mix}) as the only source of kinetic mixing.  After diagonalizing the kinetic terms, the $Z'$ couples to charged particles as a photon would, but with a coupling suppressed by $\eps$. In the limits
$m_{\chi_2}-m_{\chi_1}\ll m_{\chi_{1,2}}$ and
$m_{\chi_2}-m_{\chi_1}\ll m_{Z'}$, the partial width for the process in Eq.~(\ref{eq:off-shell-process}) can be estimated as
\bea
\Gamma(\chi_2\to \chi_1 \mu^+ \mu^-) \sim \frac{g_\chi^2 \eps^2 \alpha \,(m_{\chi_2}-m_{\chi_1})^5}
{15 \pi^2 \,m_{Z'}^4},
\label{eq:invdecaywidth}
\eea
corresponding to the partial lifetime
\bea
&&\tau_{\chi_2 \rightarrow \chi_1 \mu^+ \mu^-} 
\sim  
 8 \times 10^{19} \, {\rm s} \nonumber \\ 
&&\times \bigg(\frac{1}{g_{\chi}}\bigg)^2 \bigg(\frac{10^{-20}}{\varepsilon}\bigg)^2 \bigg(\frac{300 \, {\rm GeV}}{m_{\chi_2}-m_{\chi_1}}\bigg)^5 \bigg(\frac{m_{Z'}}{1 \, {\rm TeV}}\bigg)^4\,.
\nonumber \\
\label{eq:lifetime}
\eea
Summing over the Standard Model final states accessible through the virtual photon-like current, we estimate that the partial width to $\mu^+\mu^-$ will constitute $\sim 10\%$ of the total $\chi_2$ decay width for the benchmark considered here (see, for example, Ref.~\cite{Gopalakrishna:2008dv}).  Hence, the total lifetime of $\chi_2$ will be roughly 10 times shorter than the above estimate.

Decays can also proceed in this model in vacuum through the process $\chi_2 \to \chi_1\, Z'^{*}\to \chi_1 \,\gamma \, \phi\,\phi$, with a width given by 
\begin{align}
\Gamma(\chi_2\to \chi_1 \gamma \phi \phi) \sim \frac{g_\chi^2  \,(m_{\chi_2}-m_{\chi_1})^9}
{120,960 \,\pi^5 \,m_{Z'}^4\, M^4}.
\end{align}
To obtain $\eps\sim 10^{-20}$
for $y_N\sim 10^{-24}$ (near the maximum value allowed) requires $M\sim 9\times 10^{13}$~GeV. For this and our other benchmark parameters, the decay width in vacuum is smaller than that to muons in the proximity of Earth by a factor of $\sim 10^9$, making it consistent with constraints from observations of the isotropic gamma-ray background~\cite{DiMauro:2015tfa,Blanco:2018esa}.

\section{Decay Kinematics}
\label{sec:kinematics}

The matter-induced decays we are considering in this paper will not necessarily lead to back-to-back muon tracks, but rather to pairs of muons emerging from a common vertex with a distribution of opening angles.

In the matter-induced mass-shift scenario described in Sec.~\ref{sec:massshifts}, the $Z'$ is produced on-shell, with energy and momentum given by
\begin{align}
E_{Z'} = \frac{(m_{\chi_2}^{\rm eff})^2-(m_{\chi_1}^{\rm eff})^2 +m^2_{Z'}}{2 m_{\chi_2}^{\rm eff}}
\end{align}
and
\begin{align}
p_{Z'} = \frac{\lambda^{1/2}[(m_{\chi_2}^{\rm eff})^2,(m_{\chi_1}^{\rm eff})^2,m_{Z'}^2]}{2 m_{\chi_2}^{\rm eff}},
\end{align}
where again $\lambda(a,b,c) = a^2+b^2+c^2-2ab-2ac-2bc$.

For the sparsely instrumented arrays of IceCube and KM3NeT/ARCA, efficient reconstruction generally requires muon energies of $E_{\mu} \sim \mathcal{O}(100 \, {\rm GeV})$ or higher. Consequently, such telescopes will be sensitive to dimuon events only if $m_{\chi_2}-m_{\chi_1} \gsim 200 \, {\rm GeV}$. Denser detectors, including Super-Kamiokande, KM3NeT/ORCA, and the IceCube Upgrade, could potentially probe scenarios featuring smaller mass splittings.

The opening angle between the two muon tracks (in the $m_{\mu} \rightarrow 0$ limit) is given by
\begin{align}
\cos \theta_{\mu} = 1- \frac{2m_{Z'}^2}{E_{Z'}^2(1-\beta_{Z'}^2 \cos^2 \theta)},
\end{align}
where $\beta_{Z'}=[1- (m_{Z'}/E_{Z'})^2]^{1/2}$ is the speed of the $Z'$ and $\theta$ is the angle, measured in the $Z'$ rest frame, between the momentum of one muon and the boost axis defined by the direction of the $Z'$ momentum in the $\chi_2$ rest frame. Near the two-body threshold, $\beta_{Z'} \approx 0$, the muons will be approximately back-to-back. As the ratio, $(m^{\rm eff}_{\chi_2}-m^{\rm eff}_{\chi_1})/m_Z'$, increases, the $Z'$ becomes more highly boosted and the minimum possible opening angle decreases. 

For $m_{\chi_2}^{\rm eff} = 1 \,{\rm TeV}$,  $m_{\chi_1}^{\rm eff} = 700 \,{\rm GeV}$, and $m_{Z'}=200 \, {\rm GeV}$, the kinematically allowed opening angles range from approximately $93.3^{\circ}$ to $180^{\circ}$. Requiring both muons to have $E_{\mu} > 100 \, {\rm GeV}$ restricts the range of opening angles to approximately $93.3^{\circ}$--\,$98.2^{\circ}$.

In the case of the matter-induced kinetic mixing, as described in Sec.~\ref{sec:couplingshifts}, the $Z'$ will be produced off-shell, resulting in a distribution of invariant masses for the muon pair, $m_{\mu \mu}$. The energies of the muons can be expressed as functions of the $\chi_{1,2}$ masses and $m_{\mu \mu}$:
\begin{align}
E_{\mu^{\mp}} = 
\frac{m_{\chi_2}^2-m^2_{\chi_1} +m^2_{\mu \mu} \pm \lambda^{1/2}(m_{\chi_2}^2,m^2_{\chi_1}, m^2_{\mu \mu})\cos \theta}{4 m_{\chi_2}},
\end{align}
where $\theta$ is the angle of the muon in the dilepton rest frame relative to the motion of the dilepton system. The opening angle between the two muons is then given by 
\begin{align}
\cos \theta_{\mu} = 1 - \frac{8 m_{\chi_2}^2 m_{\mu \mu}^2}{(m_{\chi_2}^2-m^2_{\chi_1}+m^2_{\mu \mu})^2-\lambda(m_{\chi_2}^2,m^2_{\chi_1}, m^2_{\mu \mu})\cos^2 \theta}.\nonumber \\
\end{align}

To determine the distribution of muon energies and opening angles, we require the joint distribution in $m_{\mu \mu}$ and $\cos \theta$:
\begin{align}
&\frac{d^2\Gamma}{dm_{\mu \mu} \, d\cos \theta} \propto \frac{\lambda^{1/2}(m_{\chi_2}^2,m^2_{\chi_1}, m^2_{\mu \mu})}{(m_{Z'}^2-m_{\mu \mu}^2)^2}
[(m_{\chi_2}-m_{\chi_1})^2-m^2_{\mu \mu}] \nonumber \\
&\,\,\,\,\,\,\times m_{\mu \mu}[(m_{\chi_2}+m_{\chi_1})^2 (1-\cos^2\theta)+m^2_{\mu \mu}(1+\cos^2\theta)].\nonumber \\
\end{align}
For small $m_{\mu \mu}$, the dimuon system carries a large boost in the $\chi_2$ rest frame, and the two muons tend to be relatively collimated. As $m_{\mu \mu}$ approaches its kinematic maximum, $m_{\chi_2}-m_{\chi_1}$, the momentum of the dimuon system approaches zero, and the two muons become back-to-back in the $\chi_2$ rest frame. For $m_{\chi_2} = 1 \, {\rm TeV}$ and $m_{\chi_1} = 700 \, {\rm GeV}$, the resulting distribution has a median opening angle of approximately $75^{\circ}$.

\section{Neutron Star Heating}
\label{sec:NS}

One might be concerned that the enhanced decay rate of $\chi_2$ into muons and other Standard Model particles could lead to the anomalous heating of neutron stars. In this section, we provide an order-of-magnitude estimate of this effect. 

In the unscreened, linear regime, the scalar field sourced by a spherical body scales approximately as $\phi \propto N/R$, where $N$ is the number of particles sourcing the field. A neutron star contains $\sim 10^{57}$ nucleons. Comparing this with that of the Earth gives $\phi_{\rm ns}\sim 10^9 \phi(R_\oplus)$. This estimate, however, should be regarded with caution. In particular, in the matter-induced mass-shift scenario, extrapolating the linear scalar profile to neutron-star densities could produce effective mass shifts far beyond the regime in which the low-energy model is applicable. 

In the matter-induced kinetic-mixing scenario, the decay width scales as $\Gamma_{\chi_2} \propto \phi^4$. Thus, in the absence of screening, the lifetime inside a neutron star could be shorter than that near the Earth by a factor of $\sim 10^{36}$. The corresponding enhancement in the mass-shift scenario is more model dependent, because the decay width is highly sensitive to the proximity of the effective mass splitting to the on-shell threshold.

For illustration, suppose that every $\chi_2$ that is incident upon a neutron star decays and deposits an order one fraction of $m_{\chi_2}-m_{\chi_1} =300 \, {\rm GeV}$ onto that star. Neglecting gravitational focusing, this would lead to a total energy deposition rate given by 
\begin{align}
\frac{dE}{dt} &\sim n_{\chi_2} \pi R_{\rm ns}^2 v_{\rm DM} E_{\rm dep} \\
&\sim 4 \times 10^{18} \, {\rm GeV \, s}^{-1} \times \bigg(\frac{f_{\chi_2}}{0.5}\bigg)  \bigg(\frac{1 \, {\rm TeV}}{m_{\chi_2}}\bigg) \bigg(\frac{E_{\rm dep}}{200 \, {\rm GeV}}\bigg), \nonumber
\end{align}
where we have adopted $\rho_{\rm DM} = 0.4 \, {\rm GeV \, cm}^{-3}$, $R_{\rm ns}=10 \, {\rm km}$, and $v_{\rm DM} =300 \, {\rm km \, s}^{-1}$.

In steady state, the energy deposition rate is balanced by blackbody emission, $dE/dt \sim 4 \pi R_{\rm ns}^2 \sigma_{\rm SB} T_{\rm ns}^4$, where $\sigma_{\rm SB}= \pi^2/60$ is the Stefan-Boltzmann constant and $T_{\rm ns}$ is the neutron star temperature. For the rate given above, we find a steady state temperature of $T_{\rm ns}\sim 5 \times 10^{-3}$~eV, which is well below current observational sensitivities~\cite{Guillot:2019ugf}~\cite{Yakovlev:2004iq}. We therefore do not expect neutron-star temperature measurements to constrain the class of scenarios considered here.

\section{CHAMPs at IceCube}
\label{sec:CHAMPs}

Another possibility that could lead to multi-muon events with energies $E_\mu \gtrsim 100$~GeV in large-volume neutrino telescopes is the decay of long-lived charged massive particles, or CHAMPs, denoted by $X^\pm$. Collider constraints on such particles depend on their production mechanism, quantum numbers, and lifetime, but generally require $m_X \gtrsim 1 \, {\rm TeV}$. In the mass range of $\text{1~TeV} \lesssim m_X \lesssim \text{1000~TeV}$, the abundance of CHAMPs is constrained to be less than $\sim 10^{-5}$ of the local dark matter density (see, {\it e.g.}, Ref.~\cite{Ebadi:2026umu}).  Nevertheless, a small population of CHAMPs could have accumulated within the Earth at various depths. Terrestrial searches for anomalously heavy particles provide particularly stringent limits on such a population. Analyses of deep-sea water samples collected several kilometers below the surface constrain the CHAMP abundance per proton at a level of $Y_X\equiv n_X/n_p \lesssim 10^{-16}$ for masses of a few TeV; the bounds are weaker for heavier CHAMPs~\cite{Yamagata:1993jq}.

A fermionic CHAMP with a mass of $m_X\sim 1$~TeV or greater could decay through a process such as
\begin{equation}
    X^\pm \to \mu^+ \mu^- f^\pm \,,
    \label{eq:X-decay}
\end{equation}
where $f^{\pm}$ is a Standard Model or beyond-the-Standard-Model fermion with an electric charge equal to that of the parent CHAMP. This decay could produce a pair of energetic, non-collinear muons resembling the matter-induced dark matter signals discussed in the preceding sections.  

A cubic kilometer of ice contains approximately $N_p\sim \ord{10^{39}}$ protons, which implies the presence of $N_X \sim Y_X N_p$ CHAMPs in the same volume.  Hence, we estimate the following rate of CHAMP decays:
\bea
\Gamma( X^\pm \to \mu^+ \mu^- f^\pm) &\sim& 10^3~\text{yr}^{-1} \, {\rm km}^{-3}\\ \nonumber 
&\times&\left(\frac{Y_X}{10^{-16}}\right)
\left(\frac{10^{20}~\text{yr}}{\tau_X} 
\right)\, \,B_{\mu}\,,
\label{eq:Xdecay-rate}
\eea
where $B_{\mu}$ is the branching fraction into final states containing a $\mu^+ \mu^-$ pair. This result suggests that large-volume neutrino telescopes could provide an interesting venue for detecting long-lived but unstable CHAMPs. We leave a more detailed analysis of this possibility to future work.

\section{Summary and Conclusions}
\label{sec:summary}

In this study, we have considered scenarios in which dark matter particles decay within large-volume neutrino telescopes, such as IceCube or KM3NeT, producing pairs of energetic, highly non-collimated muon tracks emerging from a common vertex. Such events would have negligible Standard Model background and would constitute a smoking-gun signature of new physics.

Constraints from the cosmic-ray positron spectrum and the isotropic gamma-ray background make it highly unlikely that conventional dark matter annihilation or decay would produce an observable rate of such dimuon events in existing or planned detectors. In light of this, we have proposed two classes of models in which an excited dark matter state is extremely long-lived in vacuum but decays much more rapidly near sufficient concentrations of visible matter. In the first of these scenarios, ordinary matter sources a long-range scalar background that modifies the dark-sector mass spectrum, allowing these decays to occur near the Earth. In the second scenario, a long-range scalar induces kinetic mixing between a new gauge boson and the photon, enhancing the decay of the excited dark matter state in matter-rich environments.

We have identified viable regions of parameter space that could yield potentially observable rates of dimuon event in existing large-volume neutrino telescopes, including IceCube~\cite{IceCube:2016zyt}, KM3NeT~\cite{KM3Net:2016zxf}, Baikal-GVD~\cite{Baikal-GVD:2018isr}, and Super-Kamionkande~\cite{Super-Kamiokande:2002weg}. The unusual topology and low expected background of these events motivate dedicated searches for matter-induced dark matter decay in current and future neutrino telescopes. We have also examined the prospects for detecting multi-muon events arising from the decays of cosmologically long-lived CHAMPs and found that large-volume neutrino telescopes could provide a powerful probe of the relevant open parameter space.

\bigskip 

{\it Acknowledgments:} 
HD is supported by the US Department of Energy under Grant Contract DE-SC0012704. DH and SJ are supported by the Office of the Vice Chancellor for Research at the University of Wisconsin–Madison with funding from the Wisconsin Alumni Research Foundation.

\bibliography{apssamp}
\end{document}